\documentclass[10pt, reprint,nolongbibliography,nofootinbib,groupedaddress,showpacs,showkeys,amsmath,amssymb,eqsecnum,aps,prd]{revtex4-2}
\usepackage[final]{microtype} 
\usepackage[dvipsnames]{xcolor} 
\usepackage{graphicx} 

\usepackage{amsmath}  
\usepackage{amsfonts, amsthm} 
\usepackage{amssymb,bm}
\usepackage{newtx}

\usepackage{physics}
\usepackage{dsfont, enumitem}

\usepackage{xparse}
\usepackage{etoolbox}
\usepackage{hyperref}
\definecolor{darkorange}{RGB}{179, 74, 0}
\definecolor{complementblue}{RGB}{30, 90, 160}
\hypersetup{
    colorlinks=true,
    linkcolor=complementblue,  
    citecolor=complementblue,          
    urlcolor=darkorange,     
    filecolor=Blue,
    pdfborder={0 0 0}
}
\preto\subequations{\ifhmode\unskip\fi}

\begin{document}

\title{Spin correlations in two-particle systems:
a pedagogically motivated comparison of computational approaches}

\begin{abstract}
In this work we present a pedagogically motivated analysis of spin-correlation calculations in a quantum system composed of two spin-$1/2$ particles. Rather than aiming at new physical results, our purpose is to clarify and bring attention to different strategies for evaluating expectation values of the form $\ev{S^{(1)}_{\hat{\bm{u}}}S^{(2)}_{\hat{\bm{v}}}}{\psi}$, which play an important role in discussions of entanglement and Bell-type correlations. We compare three complementary approaches. The first follows a direct algebraic evaluation in the product basis, closely related to standard textbook methods. The second uses a matrix representation of bipartite states, in which the tensor-product structure is expressed in terms of $2\times2$ complex matrices. This representation keeps the calculation close to the familiar Pauli-matrix algebra and makes the independent action of operators on each subsystem more transparent. The third explores a symmetry-based argument, highlighting both its usefulness and its limitations when applied beyond the singlet state. We show explicitly that the singlet state is rotationally invariant, which explains why the symmetry argument successfully reproduces its correlation function, while a naive extension fails for triplet states. The discussion illustrates how entanglement, tensor-product structure, and rotational symmetry interplay in spin correlations.
\end{abstract}
\keywords{Bipartite system, spin, quantum mechanics, representation theorem}

\author{S. Martins-Filho}   
\email{s.martins-filho@unesp.br}
\affiliation{Instituto de F\'{\i}sica Te\'orica, Universidade Estadual Paulista (UNESP), Rua Dr.\ Bento Teobaldo Ferraz 271 - Bloco II, 01140-070 S\~ao Paulo, SP, Brazil}
\maketitle 

\section{Introduction}
The aim of this work is to present a pedagogically motivated discussion of
spin-correlation calculations in two-particle quantum systems, focusing on
methods suitable for advanced undergraduate or introductory graduate courses in
quantum mechanics. Rather than introducing new physical results, the goal is to
compare different derivational approaches and clarify their conceptual
structure, highlighting aspects that are often implicit in standard
presentations.

A recurring difficulty in the teaching of bipartite spin systems is that
students may be able to manipulate tensor-product expressions formally, while
still lacking a clear physical interpretation of the composite state space and
the associated correlations. In particular, the transition from single-particle
spin states to the tensor-product structure of two-particle systems is often
presented in an abstract way, which may obscure the connection between
algebraic expressions and measurable quantities.

In this work, we aim to bridge this gap by providing a unified discussion of
different computational strategies for evaluating expectation values of the
form $\ev{S^{(1)}_{\hat{\bm{u}}} S^{(2)}_{\hat{\bm{v}}}}{\psi}$, which arise naturally in
the analysis of spin correlations and Bell-type experiments. By presenting
multiple approaches within the same framework, we seek to make explicit the
relation between formal manipulations, geometric symmetry, and physical
interpretation.

We compare three complementary approaches. The first follows a direct
algebraic evaluation in the product basis, closely related to standard textbook
treatments (see, e.g., Refs.~\cite{nielsen, sakurai, cohen, ballentine}). The
second is based on a representation in which bipartite states are written as
$2\times2$ complex matrices. This choice is pedagogically useful because
students usually encounter spin-$1/2$ first through the Pauli matrices, so the
calculation remains close to familiar $2\times2$ matrix operations instead of
immediately moving to a more cumbersome $4\times4$ matrix representation. The third
explores a symmetry-based argument inspired by Griffiths \cite{G1}, highlighting both
its usefulness and its limitations when applied to the triplet sector.

The pedagogical value of the comparison is that each method emphasizes a
different aspect of the problem. The direct calculation keeps the connection
with the product basis explicit; the matrix representation makes the
subsystem structure more transparent; and the symmetry-based discussion
clarifies why rotational invariance is sufficient for the singlet but not for
the triplet sector. In particular, we analyze a common source of confusion:
the implicit assumption that symmetry arguments can be applied uniformly to
all total-spin states. By showing explicitly why this reasoning succeeds for
the singlet but fails for the triplet states, we provide a concrete example of
how symmetry considerations must be applied with care.

The target reader is an advanced undergraduate or beginning graduate student
who has already encountered Pauli matrices, Dirac notation, and the basic
postulates of quantum mechanics, but who may still be developing intuition for
composite Hilbert spaces and entangled spin states. The purpose of the
alternative derivations presented here is therefore not to replace standard
textbook methods, but to address specific difficulties that commonly arise at
this level: the interpretation of tensor-product states, the organization of
lengthy algebraic calculations, and the distinction between symmetry arguments
that rely on rotational invariance and those that do not.

This paper is organized as follows. In Sec.~\ref{sec:system} we introduce the
bipartite spin-$1/2$ system, establish the notation used throughout the paper,
and provide a physical motivation for the construction of the state space,
including its connection with spin-correlation experiments. 
In Sec.~\ref{sec:the_problem}, we formulate the spin-correlation problem.
Section~\ref{sec:methods} presents the product-basis calculation and
the matrix representation of bipartite states, allowing a direct comparison
between the two derivations. In Sec.~V, we discuss the symmetry-based approach
and its limitations when applied beyond the singlet state. Finally,
Sec.~VI summarizes the main results and discusses their pedagogical
implications.

\section{Bipartite spin system} \label{sec:system}

\begin{figure}[ht]
    \centering
    \includegraphics[width=0.3\textwidth]{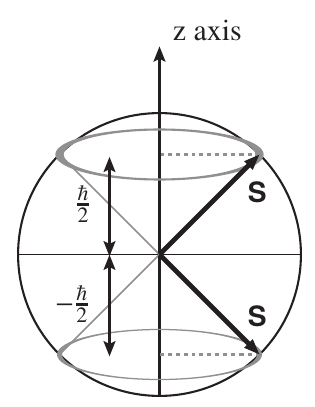}
\caption{The spin-$1/2$ degrees of freedom. For a measurement along the $z$ axis, the possible outcomes are only $S_z=+\hbar/2$ and $S_z=-\hbar/2$. The arrows represent a visual aid and should not be interpreted as classical spin vectors.}
    \label{fig:spin12}
\end{figure}

Before introducing the formal Hilbert-space description, it is interesting to recall the physical meaning of spin-$1/2$ systems and the construction of bipartite states from a more intuitive perspective. A spin-$1/2$ particle is characterized by the fact that any measurement of its spin component along a given direction yields only two possible outcomes, $+\hbar/2$ or $-\hbar/2$ (see Fig.~\ref{fig:spin12}). These two outcomes are associated with the two eigenstates of the spin operator along the chosen axis. Thus, even before introducing matrices, the essential physical content is that a spin-$1/2$ system behaves as a two-outcome quantum system for each chosen measurement direction.

\begin{figure*}[ht]
    \centering
    \includegraphics[width=0.95\textwidth]{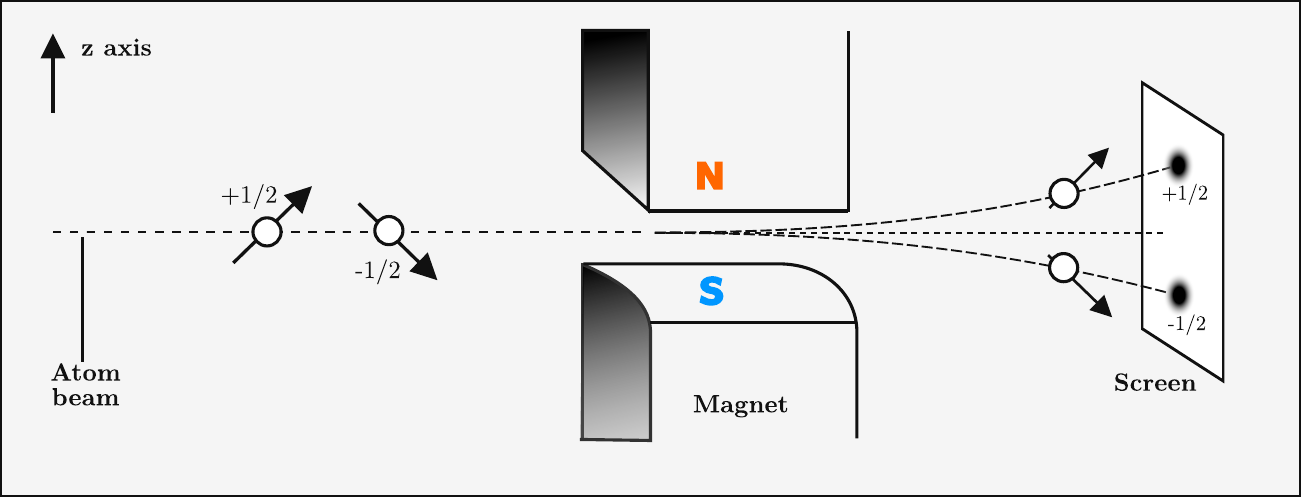}
\caption{Schematic representation of a Stern--Gerlach measurement for a
spin-$1/2$ system. 
The inhomogeneous magnetic field produced by the
magnet separates particles according to the measured value of the spin
component along a selected direction.  In this
case, only two outcomes are possible, $+\hbar/2$ and $-\hbar/2$, corresponding
respectively to spin ``up'' and spin ``down'' along the chosen axis.}
    \label{fig:ST}
\end{figure*}

In practice, such measurements can be realized, for example, with a Stern--Gerlach apparatus \cite{Stern, GandS}, which separates a beam of particles according to the value of the spin component along a chosen axis, as depicted in Fig.~\ref{fig:ST}. For readers interested in the historical impact of this experiment, see Ref.~\cite{history, his2, Grossi2023}. The quantum state does not assign a pre-existing classical direction to the spin; rather, it encodes the probability amplitudes associated with the possible measurement outcomes.

When two spin-$1/2$ particles are considered, the physical description must include all possible joint outcomes of measurements performed on the two particles. If the spin of each particle is measured along the same quantization axis, there are four possible joint results:
\[
(+,+), \quad (+,-), \quad (-,+), \quad (-,-).
\]
These outcomes correspond respectively to the product states
\begin{equation*}
    \begin{split}
& \ket{+\frac{1}{2}}^{(1)}\ket{+\frac{1}{2}}^{(2)}, \quad
\ket{+\frac{1}{2}}^{(1)}\ket{-\frac{1}{2}}^{(2)}, \\
& \ket{-\frac{1}{2}}^{(1)}\ket{+\frac{1}{2}}^{(2)}, \quad
\ket{-\frac{1}{2}}^{(1)}\ket{-\frac{1}{2}}^{(2)};
\end{split}
\end{equation*}
where the superscripts $(1)$ and $(2)$ indicate the particle to which each state refers.
Thus, the tensor-product structure $\mathcal{H}\otimes\mathcal{H}$ can be understood physically as the state space generated by all possible superpositions of joint measurement outcomes. In this sense, the tensor product is not only a formal mathematical construction, but also the natural way of describing the outcomes of measurements performed on a composite quantum system.

A particularly important feature of bipartite quantum systems is the existence
of entangled states, which cannot be written as a simple product of
single-particle states. In the two-spin system, examples of such states are
superpositions of the joint outcomes $(+,-)$ and $(-,+)$, in which the two
particles do not possess independently assigned spin values before measurement.
These states exhibit correlations that cannot be interpreted as a mere lack of
knowledge about pre-existing classical values. 
The formal
examples of this idea, namely the singlet and triplet states, will be introduced
in the next subsection after the notation for the bipartite Hilbert space has
been established.

The central quantity studied in this work is the correlation function
\begin{equation}
    B \left [ {\psi}\right ] \equiv
    \ev**{S^{(1)}_{\hat{\bm{u}}} S^{(2)}_{\hat{\bm{v}}}}{\psi},
    \label{eq:bellexp}
\end{equation}
where $\ket{\psi}\in \mathcal{H}\otimes\mathcal{H}$ is a state of the 
bipartite system. This quantity has a direct physical interpretation: it is the
average product of spin measurements performed on particles 1 and 2 along the
directions $\hat{\bm{u}}$ and $\hat{\bm{v}}$, respectively, when the system is
prepared in the state $\ket{\psi}$.

In a Stern-Gerlach experiment, each individual measurement gives either $+\hbar/2$ or $-\hbar/2$, and the correlation function measures how the two outcomes are statistically related when the experiment is repeated many times under the same preparation. 
Such correlations are central in discussions of quantum entanglement and
Bell-type experiments \cite{bell, clauser}. Pedagogical discussions of Bell
inequalities and their conceptual implications can be found, for example, in
Refs.~\cite{Machado2023,Pimentel2025}.

From a pedagogical perspective, a common difficulty for students is to interpret the tensor-product structure purely as an abstract construction, without connecting it to measurable quantities. By framing bipartite states in terms of joint measurement outcomes, the present discussion aims to make this connection explicit and to provide a more intuitive understanding of how spin correlations arise in composite systems. This perspective also helps clarify potential misconceptions, such as the improper extension of symmetry arguments to states that are not rotationally invariant.

This physical picture provides the basis for the formal developments presented in the following sections. The different computational approaches discussed below should therefore be understood not merely as algebraic alternatives, but as complementary ways of organizing the same physical information: the relation between the preparation of a two-particle spin state and the correlations observed in joint spin measurements.

\subsection{Formal Hilbert space}
To proceed, we establish the notation that will be used throughout the paper.
Let us denote, respectively, $ \bm{S}^{(1)} $ and $ \bm{S}^{(2)} $ as the spin operators of particles 1 and 2. Formally, the system space $ \hat{\mathcal{H}} $ is given by the tensor product $ \mathcal{H} \otimes \mathcal{H}$, where $ \mathcal{H} $ is a two-dimensional complex Hilbert space \cite{sakurai,cohen,nielsen}. Thus, $ \bm{S}^{(1)} \equiv \bm{\mathsf{S}} \otimes \mathds{1} $ and $ \bm{S}^{(2)} \equiv \mathds{1} \otimes \bm{\mathsf{S}} $. Note that $ \bm{\mathsf{S}} \equiv \hbar \bm{\sigma}/2$ is the well-known spin operator of a single spin-$1/2$ particle, and $ \bm{\sigma} $ is the Pauli vector defined by the following components\footnote{The components are known as the Pauli matrices \cite{ballentine}. They are conveniently labeled by numbers since the Einstein summation convention will be used throughout.}
\begin{equation}
\sigma_1=\begin{pmatrix}\pmat{1} \end{pmatrix}, \quad
\sigma_2=\begin{pmatrix}\pmat{2} \end{pmatrix},\quad \textrm{and} \quad
\sigma_3=\begin{pmatrix}\pmat{3} \end{pmatrix}.
\end{equation}

In general, the operator $ A^{(1)} \equiv A \otimes \mathds{1} $ while $ A^{(2)} \equiv \mathds{1} \otimes A$, where $ A $ belongs to the set of all linear operators on $ \mathcal{H} $, denoted by $ \mathcal{L} ( \mathcal{H} ) $. Clearly, when $ A \in \mathcal{L} ( \mathcal{H} )$, the operator $ \hat{A} \equiv A \otimes A \in \mathcal{L} ( \mathcal{H} \otimes \mathcal{H} ) $.

The components of $ \bm{\mathsf{S}} $ satisfy the $ \mathfrak{su} (2) $ algebra, that is, \cite{edmonds}
\begin{equation}
    [ {\mathsf{S}}_{j} ,\, {\mathsf{S}}_k]=i \hbar \epsilon_{jkl} {\mathsf{S}}_l,
    \label{eq:spinC}
\end{equation}
where the Einstein summation convention, with indices running from $ 1 $ to $ 3 $, is used and will be adopted throughout the text.

The eigenstates of ${\mathsf{S}}_{z}$ are
\begin{equation}
    \ket{\frac{1}{2}}\equiv \chi_+
    =
    \begin{pmatrix}
        1 \\ 0
    \end{pmatrix},
    \quad \text{and} \quad
    \ket{-\frac{1}{2}}\equiv \chi_-
    =
    \begin{pmatrix}
        0 \\ 1
    \end{pmatrix},
    \label{eq:estates}
\end{equation}
with
\begin{equation}
    {\mathsf{S}}_{z}\chi_{\pm}
    =
    \pm \frac{\hbar}{2}\chi_{\pm}.
\end{equation}
They form a basis for $\mathcal{H}$, while their tensor products form a basis
for $\hat{\mathcal{H}}$, commonly referred to as the \emph{product basis}.

Henceforth, we denote the total spin operator $ \bm{S}^{(1)} + \bm{S}^{(2)} $ simply by $ \bm{S} $. That is,
\begin{equation}
    \bm{S} \equiv   \bm{S}^{(1)} + \bm{S}^{(2)} = \frac{\hbar}{2} \left ( 
        \bm{\sigma} \otimes \mathds{1} + \mathds{1} \otimes \bm{\sigma} 
    \right ).
\end{equation}

We are particularly interested in the total-spin basis formed by the simultaneous eigenstates of $ S_{z} \equiv S_z^{(1)} + S_{z}^{(2)} $ and $ \bm{S}^{2} $. These eigenstates consist of the singlet $\ket{0}$, with total spin $s=0$, and the triplet states $\ket{1,1}$, $\ket{1,0}$, and $\ket{1,-1}$ with $s=1$. In this basis, the singlet satisfies
\begin{equation}
S_z \ket{0} = S^2 \ket{0} = 0\ket{0},
\end{equation}
while the triplet states satisfy
\begin{align}
    S_z \ket{1,1} &= \hbar \ket{1,1},\\
    S_z \ket{1,0} &= 0 \ket{1,0}, \quad \textrm{and} \\
    S_z \ket{1,-1} &= -\hbar \ket{1,-1}.
\end{align}
Thus, the possible measurement outcomes of the total spin component $S_z$ are $+\hbar$, $0$, and $-\hbar$, respectively.

If we generalize the notation such that
$ A\otimes B \equiv A^{(1)}B^{(2)}$, the eigenstates in the standard basis can be written as
\begin{subequations} \allowdisplaybreaks
\label{eq:states}
    \begin{align}
        \ket{1,1} &= \chi_+^{(1)}\chi_+^{(2)}, \\
        \ket{1,-1} &= \chi_-^{(1)}\chi_-^{(2)}, \\
        \ket{1,0} &= \frac{\chi_{+}^{(1)} \chi_{-}^{(2)}+\chi_{-}^{(1)}\chi_{+}^{(2)}}{\sqrt{2}}, \quad \textrm{and}\\
        \ket{0} &= \frac{\chi_{+}^{(1)} \chi_{-}^{(2)}-\chi_{-}^{(1)}\chi_{+}^{(2)}}{\sqrt{2}}.
    \end{align}
\end{subequations}

This notation is particularly convenient when decomposing operators into the
two independent sectors associated with particles 1 and 2. If
$\hat{C}=A^{(1)}B^{(2)}$, with $C^{(1)}=A\otimes\mathds{1}$ and
$A^{(2)}=\mathds{1}\otimes B$, then the two operators commute,
\begin{equation}
    [A^{(1)},B^{(2)}]=0,
\end{equation}
(even when $[A, B] \neq 0$)
because they act on different factors of the tensor-product space. More
explicitly, for a product state $\chi_{\alpha}^{(1)}\chi_{\beta}^{(2)}$, with
$\alpha,\beta=\pm$, one has
\begin{equation}
A^{(1)}B^{(2)}
\chi_{\alpha}^{(1)}\chi_{\beta}^{(2)}
=
(A\chi_{\alpha}^{(1)})
(B\chi_{\beta}^{(2)}).
\label{eq:c1}
\end{equation}
Thus, for example, the operator $\hat{C}$ acts on the singlet state as
\begin{equation}
A^{(1)}B^{(2)} \ket{0}
=
\frac{
(A\chi_+^{(1)})(B\chi_-^{(2)})
-
(A\chi_-^{(1)})(B\chi_+^{(2)})
}{\sqrt{2}}.
\label{eq:c2}
\end{equation}

\section{Statement of the problem}\label{sec:the_problem}

\begin{figure}[ht]
    \centering
    \includegraphics[width=0.40\textwidth]{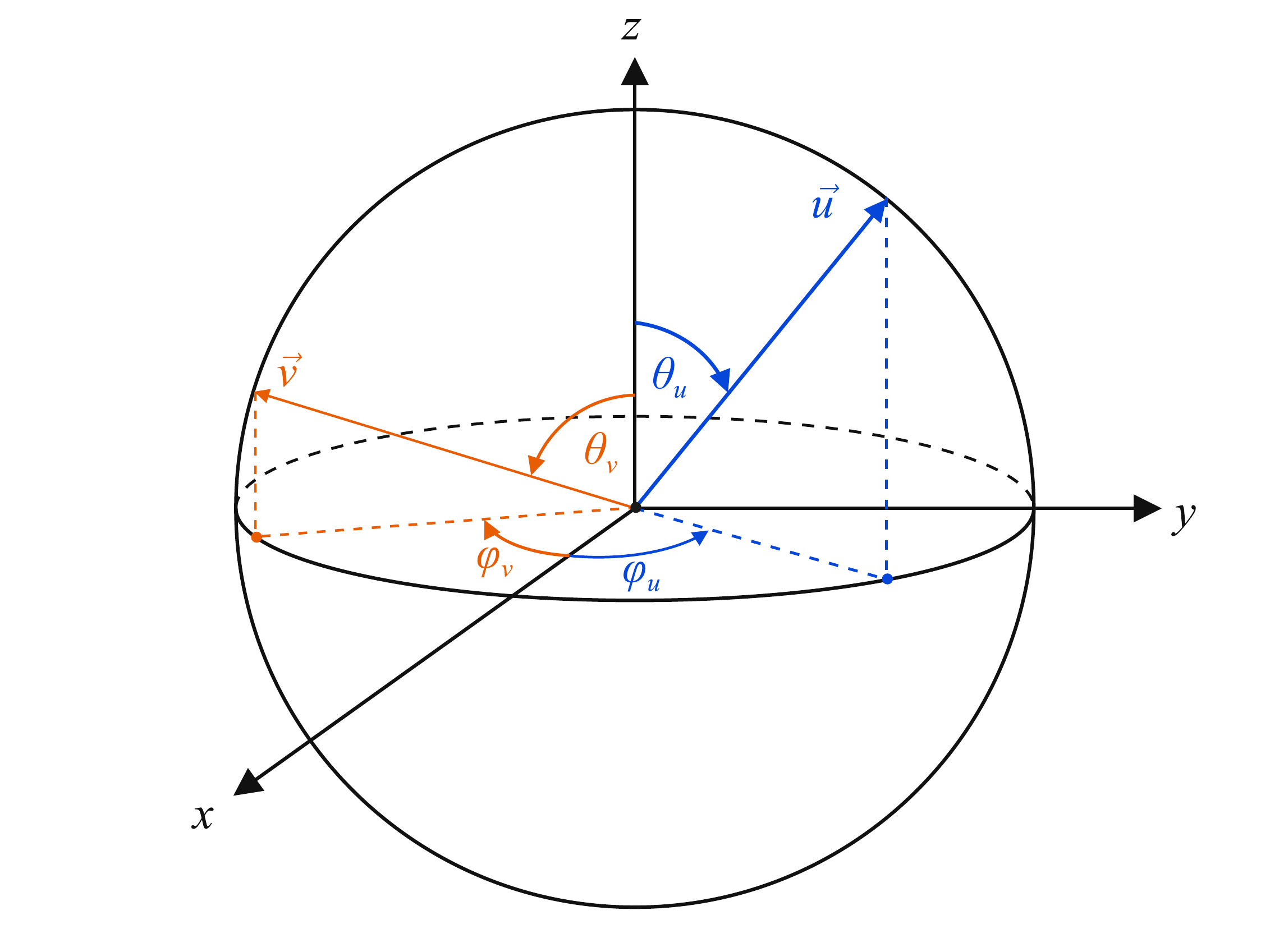}
    \caption{Geometrical representation of the measurement directions $\hat{\bm{u}}$ and $\hat{\bm{v}}$ in a unit sphere ($r^{2} =1$) using spherical coordinates. The polar angles $\theta_u$ and $\theta_v$ are measured from the positive $z$-axis, while the azimuthal angles $\varphi_u$ and $\varphi_v$ are measured in the $xy$-plane from the positive $x$-axis to the projections of $\hat{\bm{u}}$ and $\hat{\bm{v}}$, respectively.}\label{fig:sandu}
\end{figure}

Let us define the spin operator along the direction $\hat{\bm u}$, denoted by $ \mathsf{S}_{\hat{\bm{u}}}$. We adopt spherical coordinates for convenience, see Fig.~\ref{fig:sandu}. 
The unit vector $\hat{\bm u}$ can be written as
\begin{equation}
{\cos{\varphi_{u}}} {\sin{\theta_{u}}}\hat{\bm{\imath}} +{\sin{\varphi_u}}{\sin{\theta_{u}}}\hat{\bm{\jmath}} + {\cos{\theta_{u}}} \hat{\bm{k}},
\end{equation}
so that the spin operator along this direction assumes, in the usual basis, the form
\begin{equation}
    \mathsf{S}_{\hat{\bm{u}}} = \bm{\mathsf{S} } \cdot \hat{\bm{u}}=\frac{\hbar}{2}
    \begin{pmatrix}
    \cos \theta_{u} & \sin \theta_{u} e^{-i\varphi_{u}} \\
    \sin \theta_{u} e^{i\varphi_{u}} & -\cos \theta_{u}
    \end{pmatrix}.
    \label{eq:Su}
\end{equation}

Our goal is to compare different approaches for evaluating the expectation value~\eqref{eq:bellexp}, where $ S^{(1)}_{\hat{\bm{u}}} \equiv \mathsf{S}_{\hat{\bm{u}}} \otimes \mathds{1}$ 
and $ S^{(2)}_{\hat{\bm{v}}} \equiv \mathds{1} \otimes \mathsf{S}_{\hat{\bm{v}}}$. 
This requires computing $B [ \psi ]$ for the states of the total-spin basis. Such quantities naturally arise in spin–correlation measurements, including discussions of Bell-type inequalities.

\section{Evaluation of the spin correlation $ B[\psi]$}\label{sec:methods}

There are several ways to evaluate $B \left [ {\psi} \right ]$ shown in equation~\eqref{eq:bellexp}, reflecting different representations of the $\mathfrak{su}(2)\otimes \mathfrak{su}(2)$ algebra. In the following sections, we present some alternative procedures. We begin with a direct and effective, although not particularly concise, method for computing $B$.

\subsection{Product-basis resolution}

Rather than working in the standard basis, it is convenient to use the eigenvectors of the operators $S_{\hat{\bm{u}}}^{(1)}$ and $S_{\hat{\bm{v}}}^{(2)}$.
Solving the eigenvalue problem associated with the matrix in equation~\eqref{eq:Su} is straightforward. One finds that
\begin{equation}
    \chi^{(\hat{\bm{u}})}_{+}= \begin{pmatrix}
        \cos{\frac{\theta_{u}}{2}} \\
        e^{i\varphi_u}\sin{\frac{\theta_u}{2}}
    \end{pmatrix}
   \quad \textrm{and}
   \quad
 \chi^{(\hat{\bm{u}})}_{-}= \begin{pmatrix}
        e^{-i\varphi_u}\sin{\frac{\theta_u}{2}}\\
        -\cos{\frac{\theta_{u}}{2}}
    \end{pmatrix}
\label{eq:evu}
\end{equation}
are eigenvectors of $S_{\hat{\bm{u}}}^{(1)}$ with eigenvalues $+\frac{\hbar}{2}$ and $-\frac{\hbar}{2}$, respectively. Since $S_{\hat{\bm{v}}}^{(2)}$ has the same structure as $S_{\hat{\bm{u}}}^{(1)}$ under the replacement $u \rightarrow v$, the corresponding eigenvectors are
\begin{equation}\label{eq:evv}
    \chi^{(\hat{\bm{v}})}_{+}= \begin{pmatrix}
        \cos{\frac{\theta_{v}}{2}} \\
        e^{i\varphi_v}\sin{\frac{\theta_v}{2}}
    \end{pmatrix} \quad \textrm{and}
   \quad \chi^{(\hat{\bm{v}})}_{-}= \begin{pmatrix}
        e^{-i\varphi_v}\sin{\frac{\theta_v}{2}}\\
        -\cos{\frac{\theta_{v}}{2}}
    \end{pmatrix},
\end{equation}
with identical eigenvalues.

The next step consists of constructing the eigenstates in this new basis. From equation.~\eqref{eq:evu} and \eqref{eq:evv}, we obtain
\begin{subequations}
    \begin{align}
        \chi_+^{(1)}&=\chi_{+}^{(\hat{\bm{u}})}\cos{\frac{\theta_{u}}{2}}+ \chi_-^{(\hat{\bm{u}})}e^{i\varphi_{u}}\sin{\frac{\theta_{u}}{2}},
        \\
        \chi_+^{(2)}&=\chi_{+}^{(\hat{\bm{v}})}\cos{\frac{\theta_{v}}{2}}+ \chi_-^{(\hat{\bm{v}})}e^{i\varphi_{v}}\sin{\frac{\theta_{v}}{2}},
\\
\chi_-^{(1)}&=\chi_{+}^{(\hat{\bm{u}})}e^{-i\varphi_{u}}\sin{\frac{\theta_{u}}{2}}-\chi_-^{(\hat{\bm{u}})}\cos{\frac{\theta_{u}}{2}}, \quad \textrm{and} \\
\chi_-^{(2)}&=\chi_{+}^{(\hat{\bm{v}})}e^{-i\varphi_{v}}\sin{\frac{\theta_{v}}{2}}-\chi_-^{(\hat{\bm{v}})}\cos{\frac{\theta_{v}}{2}}.
    \end{align}
\label{eq:basis}
\end{subequations}

Substituting equation.~\eqref{eq:basis} into the standard-basis eigenstates given in equation~\eqref{eq:states}, we obtain that
\begin{equation}\label{eq:singletNB}
    \begin{split}
        \ket{0} = \frac{1}{\sqrt{2}} \big ( & c_{11} \chi^{(\hat{\bm{u}})}_{+}
    \chi^{(\hat{\bm{v}})}_{+} + c_{12} \chi^{(\hat{\bm{u}})}_{+} \chi^{(\hat{\bm{v}})}_{-}
                \\ & 
    - \bar{c}_{12} \chi^{(\hat{\bm{u}})}_{-} \chi^{(\hat{\bm{v}})}_{+} -
\bar{c}_{11}  \chi^{(\hat{\bm{u}})}_{-} \chi^{(\hat{\bm{v}})}_{-} \big ),
\end{split}
\end{equation}
with
\begin{subequations}
    \begin{align}
        c_{11} &=  e^{-i \varphi_v}
        \cos{\frac{\theta_u}{2}} \sin{\frac{\theta_v}{2}} -
        e^{-i \varphi_u} \sin{\frac{\theta_u}{2}} \cos{\frac{\theta_v}{2}}
        ,\intertext{and} 
        c_{12} &= -\cos{\frac{\theta_u}{2}} \cos{\frac{\theta_v}{2}} -
        e^{-i(\varphi_u-\varphi_v)} \sin{\frac{\theta_u}{2}}
        \sin{\frac{\theta_v}{2}}.
    \end{align}
\end{subequations}
While the triplet is given by:
\begin{subequations}\allowdisplaybreaks
    \begin{align}
        \ket{1,1} &=  c_{21} \chi^{(\hat{\bm{u}})}_{+}
    \chi^{(\hat{\bm{v}})}_{+} + c_{22} \chi^{(\hat{\bm{u}})}_{+} \chi^{(\hat{\bm{v}})}_{-}
        \nonumber\\ &  +  {c}_{23} \chi^{(\hat{\bm{u}})}_{-} \chi^{(\hat{\bm{v}})}_{+} +
{c}_{24}  \chi^{(\hat{\bm{u}})}_{-} \chi^{(\hat{\bm{v}})}_{-}, \\
    \ket{1,-1} &=  \bar{c}_{24} \chi^{(\hat{\bm{u}})}_{+}
    \chi^{(\hat{\bm{v}})}_{+} - \bar{c}_{23} \chi^{(\hat{\bm{u}})}_{+} \chi^{(\hat{\bm{v}})}_{-}
\nonumber\\ &
    -  \bar{c}_{22} \chi^{(\hat{\bm{u}})}_{-} \chi^{(\hat{\bm{v}})}_{+} +
    {c}_{21}  \chi^{(\hat{\bm{u}})}_{-} \chi^{(\hat{\bm{v}})}_{-}, \quad \text{and} \\
    \ket{1,0} &= c_{41} \chi^{(\hat{\bm{u}})}_{+}
    \chi^{(\hat{\bm{v}})}_{+} + c_{42} \chi^{(\hat{\bm{u}})}_{+} \chi^{(\hat{\bm{v}})}_{-}
\nonumber\\ &
    -  \bar{c}_{42} \chi^{(\hat{\bm{u}})}_{-} \chi^{(\hat{\bm{v}})}_{+} -
\bar{c}_{41}  \chi^{(\hat{\bm{u}})}_{-} \chi^{(\hat{\bm{v}})}_{-};
    \end{align}
\end{subequations}
where
\begin{subequations}
    \begin{align}
        c_{21} ={}&\cos{\frac{\theta_u}{2}} \cos{\frac{\theta_v}{2}},
        \\
        c_{22} ={}& e^{-i\varphi_v} \cos{\frac{\theta_u}{2}}
        \sin{\frac{\theta_v}{2}}, \\
        c_{23} ={}& e^{i \varphi_u} \sin{\frac{\theta_u}{2}} \cos{\frac{\theta_v}{2}
        }, \\ 
        c_{24} ={}& e^{i(\varphi_u + \varphi_v)}
        \sin{\frac{\theta_u}{2}} \sin{\frac{\theta_v}{2}};\\
        c_{41} ={}&  e^{-i \varphi_v}
        \cos{\frac{\theta_u}{2}} \sin{\frac{\theta_v}{2}} +
        e^{-i \varphi_u} \sin{\frac{\theta_u}{2}} \cos{\frac{\theta_v}{2}}
        , \\
        c_{42} ={}& -\cos{\frac{\theta_u}{2}} \cos{\frac{\theta_v}{2}} +
        e^{-i(\varphi_u-\varphi_v)} \sin{\frac{\theta_u}{2}}
        \sin{\frac{\theta_v}{2}}.
    \end{align}
\end{subequations}

The action of $S^{(1)}_{\hat{\bm{u}}} S^{(2)}_{\hat{\bm{v}}}$ on these eigenstates follows directly from equation.~\eqref{eq:c1} (with $A=B=\mathsf{S}$). Since
\begin{equation}
    S^{(1)}_{\hat{\bm{u}}} \chi_\pm^{(\hat{\bm{u}})} = \pm\frac{\hbar}{2}\chi_\pm^{(\hat{\bm{u}})}
    \quad \textrm{and} \quad S^{(2)}_{\hat{\bm{v}}} \chi_\pm^{(\hat{\bm{v}})} = \pm \frac{\hbar}{2}\chi_\pm^{(\hat{\bm{v}})},
\end{equation}
it implies  that
\begin{subequations} \label{eq:bells}
    \begin{align}
        \ev**{S^{(1)}_{\hat{\bm{u}}} S^{(2)}_{\hat{\bm{v}}}}{0}&=-\frac{\hbar^2}{4}
        \cos\theta,
        \label{eq:bellS0}\\
        \ev**{S^{(1)}_{\hat{\bm{u}}} S^{(2)}_{\hat{\bm{v}}}}{1, 0}&=
        \frac{\hbar^2}{4}  [
        \cos{\theta} -2 \cos{\theta_{u}} \cos{\theta_{v}}] ,
        \label{eq:bellSP}\\
        \ev**{S^{(1)}_{\hat{\bm{u}}} S^{(2)}_{\hat{\bm{v}}}}{1,1}&=\frac{\hbar^2}{4}\cos\theta_u
        \cos \theta_v,\quad \textrm{and}
        \label{eq:bellMR} \\
        \ev**{S^{(1)}_{\hat{\bm{u}}} S^{(2)}_{\hat{\bm{v}}}}{1,-1}&=\frac{\hbar^2}{4}\cos\theta_u
        \cos \theta_v;
        \label{eq:bellMN}
    \end{align}
\end{subequations}
where $\theta$ is the angle between $\hat{\bm{u}}$ and $\hat{\bm{v}}$, such that
\begin{align}\label{eq:angletheta}
    \cos \theta &= \hat{\bm{u}}\cdot \hat{\bm{v}}  \nonumber \\
    &=\cos \theta_u \cos \theta_v
    +\sin \theta_u \sin \theta_v \cos{(\varphi_u-\varphi_v)}.
\end{align}

Since the four total-spin eigenstates form a complete basis of 
$\hat{\mathcal{H}}$, their equal-weight sum corresponds to the maximally 
mixed state $ \mathds{1}/4$, on which the correlation vanishes 
because each single-particle spin operator is traceless. Hence, for any 
measurement directions $\hat{\boldsymbol{u}}$ and $\hat{\boldsymbol{v}}$,
\begin{equation}
B[0]+B[1,1]+B[1,-1]+B[1,0]=0,
\label{eq:sumrule}
\end{equation}
which is satisfied by our results~\eqref{eq:bells}.
This provides a consistency check on the four correlation functions.

This approach provides a direct solution to the problem. All expectation values
are obtained explicitly, and the final expressions are compact. However, the
derivation itself is rather lengthy, especially for $B \left [ {0} \right]$ and
$B \left [ 1,0 \right ]$. In these cases, several intermediate algebraic steps
are needed before the simple structure of the result becomes apparent. From a
pedagogical point of view, this is a limitation: the calculation is completely
correct, but the amount of algebra may obscure the tensor-product structure of
the problem and the physical origin of the correlations.

This motivates the search for a shorter and more organized derivation. The
purpose is not to replace the direct calculation, which remains useful as a
first explicit approach, but to complement it with a method that makes the
bipartite structure more transparent and reduces the possibility of algebraic
errors. This comparison also helps to identify which parts of the calculation
are essential to the physics and which are merely consequences of the chosen
representation.

\subsection{Matrix representation approach}\label{sec:second}
As stated at the beginning of Section~\ref{sec:methods}, there are several ways to obtain the previous result. In this section we present a more concise method for evaluating the expression~\eqref{eq:bellexp}. The approach is based on a matrix representation of the tensor-product structure, allowing bipartite states to be expressed in terms of complex matrices.

Henceforth we denote a general spinor in $\mathcal{H}$ by $\ket{a}$, where $a$ is any lowercase Latin letter, possibly primed when necessary. Its components are $\ket{a}_{1}=a_1$ and $\ket{a}_2=a_2$. The composite state $\ket{a}\otimes \ket{b}\in \mathcal{\hat H}$ will be referred to as a \emph{pure c-state} (where “c-state’’ stands for composite state). A linear superposition of pure c-states, such as the singlet, will be referred to as a \emph{mixed c-state}.

The c-state $\ket{a} \otimes \ket{b}$ is represented as the $2\times 2$ complex matrix 
\begin{equation}
    \Sigma^{ab}=\ket{a}\bra{b}^*,\label{eq:def}
\end{equation}
whose elements in the standard basis are
\begin{equation}
    (\Sigma^{ab}){}_{jk}=a_{j}b_k, \quad
    \text{with } j\text{, }k \in \{1,\, 2\}.
    \label{eq:defE}
\end{equation}
Note that $A^*$ denotes the complex conjugate of the matrix $A$, while $\bar{a}$ denotes the complex conjugate of the complex number $a$. In addition, $A^\dagger$ denotes the Hermitian transpose of $A$.

 It is important to distinguish the tensor product $\ket{a}\otimes\ket{b}$,
which defines a state in a composite Hilbert space, from the outer product
$\ket{a}\bra{b}$, which defines a linear operator acting on a single Hilbert space.
In the present work, we make use of an isomorphism that allows the tensor-product
structure to be represented in terms of $2\times2$ complex matrices, 
thereby representing composite states by matrices with outer-product-like components.

The standard basis of the $4$-dimensional Hilbert space $\mathcal{\hat H}$ defined through this product coincides with the standard basis of the vector space $\mathbb{C}^{2\times 2}$. From the definition~\eqref{eq:def} this can be verified directly:
\begin{subequations}\allowdisplaybreaks
\begin{align}
    \Sigma^{++}&=\chi_+\chi_+^\mathsf{T} =
    \begin{pmatrix}
        1 & 0 \\ 0 & 0
    \end{pmatrix}, \\
    \Sigma^{+-}&=\chi_+\chi_-^\mathsf{T} =
    \begin{pmatrix}
        0 & 1 \\ 0 & 0
    \end{pmatrix}, \\
    \Sigma^{-+}&=\chi_-\chi_+^\mathsf{T} =
    \begin{pmatrix}
        0 & 0 \\ 1 & 0
    \end{pmatrix}, \quad \text{and}\\
    \Sigma^{--}&=\chi_-\chi_-^\mathsf{T} =
    \begin{pmatrix}
        0 & 0 \\ 0 & 1
    \end{pmatrix}.
\end{align}\label{eq:Sbasis}
\end{subequations}
These matrices clearly form the standard basis of $\mathbb{C}^{2\times2}$. In fact, any basis of $\mathbb{C}^{2\times2}$ also constitutes a basis for our four-dimensional Hilbert space. Therefore, $\mathcal{\hat H}$ is isomorphic to $\mathbb{C}^{2\times2}$.

Before performing any calculations, it is useful to understand how operators act within this formulation. To this end, we recall the defining property of the tensor product. The operator $P\otimes Q$ acts independently on a pure c-state $\ket{a}\otimes \ket{b}$, namely
\begin{equation}
    \left ( P\otimes Q \right ) \ket{a}\otimes \ket{b} =
    \left ( P\ket{a} \right ) \otimes \left ( Q \ket{b}\right ).
    \label{eq:p1}
\end{equation}
We take this property as the starting point to define the action of $P\otimes Q$ in the matrix representation. By bilinearity of the tensor product, this definition extends naturally to mixed c-states. Substituting equation~\eqref{eq:def} into equation~\eqref{eq:p1} yields
\begin{equation}
    \left (P \otimes Q \right )\Sigma^{ab} = P \Sigma^{ab} Q^\mathsf{T},
    \label{eq:main}
\end{equation}

\begin{proof}
    From equation~\eqref{eq:defE}, the right-hand side of equation~\eqref{eq:p1} is a matrix whose elements are
    \begin{equation}
        \left [\left ( P\ket{a} \right ) \otimes \left ( Q \ket{b}\right )
        \right ]{}_{jk} =
        (P\ket{a}){}_j(Q\ket{b}){}_k= P_{jl}a_{l}Q_{km}b_m.
    \end{equation}
    Rearranging this expression we obtain
    \begin{equation}
        P_{jl}a_{l}b_{m}Q_{km}=P_{jl} (\Sigma^{ab}  ){}_{lm}
         ( Q^\mathsf{T}  ){}_{mk} =
         (P\Sigma^{ab} Q^\mathsf{T}  ){}_{jk}.
        \qedhere
    \end{equation}
\end{proof}

The inner product between two pure c-states is defined as
\begin{equation}
    \left \langle \Sigma^{a'b'}, \Sigma^{ab}\right \rangle
    \equiv \Tr \!\left[  ( \Sigma^{a'b'}  )^\dagger \Sigma^{ab} \right].
    \label{eq:sprod}
\end{equation}
This definition extends straightforwardly to mixed c-states. When $\ket{a'}=\ket{a}$ and $\ket{b'}=\ket{b}$, it reduces to the squared norm of $\ket{a}\otimes \ket{b}$:
\begin{equation}
    \big{\|} \ket{a}\otimes\ket{b} \big{\|}^2
    \equiv \Tr \!\left[  ( \Sigma^{ab}  )^\dagger \Sigma^{ab} \right],
\end{equation}
which corresponds to the squared norm of the tensor product $\ket{a} \otimes \ket{b}$ in this matrix representation.

Taking into account the isomorphism mentioned above, this definition naturally suggests itself as the inner product in $\mathcal{\hat H}$. Indeed, it coincides with the standard inner product, as we show below.
\begin{proof}
    The inner product in $\mathcal{\hat H}$ is
    \begin{equation}
        \left < \ket*{a'}\otimes \ket*{{b}{'}} , \, \ket{a}\otimes\ket{b}
        \right >
        \equiv \braket*{a{'}}{a} \braket*{b'}{b}.
        \label{eq:rSP}
    \end{equation}

    This agrees with
    \begin{align}
         \bra*{a'}\otimes \bra*{{b}{'}}
         \big (\ket{a}\otimes\ket{b}\big ) &\equiv
         \big ( \ket*{a'} \otimes \ket*{{b}{'}}\big )^\dagger
        \big (\ket{a}\otimes\ket{b}\big ) \nonumber \\
        &= \braket*{a{'}}{a} \braket*{b'}{b}.
    \end{align}

    Using the definition of the trace we write
    \begin{equation}
        \Tr [  ( \Sigma^{a'b'}  )^\dagger \Sigma^{ab} ] =
         ( \Sigma^{a'b'}  )^{*}_{ji} (\Sigma^{ab} )
        {}_{ji}.
    \end{equation}
    Substituting the matrix elements from equation~\eqref{eq:defE} we obtain
    \begin{align}
        \Tr [  ( \Sigma^{a'b'}  )^\dagger \Sigma^{ab} ] &=
        \bar{a'}_{j}\bar{b'}_{i}a_{j}b_i \nonumber \\
        &= \bar{a'}_{j}a_{j} \bar{b'}_{i}b_{i} \nonumber \\
        &= \braket*{a'}{a}\braket*{b'}{b}.
    \end{align}
    Therefore equation~\eqref{eq:sprod} is equivalent to the standard inner product of $\mathcal{\hat H}$ defined in equation~\eqref{eq:rSP}.
\end{proof}

With equation~\eqref{eq:main} and the inner product~\eqref{eq:sprod} we are ready to address the problem. We start by using this formalism with the standard basis defined in equation~\eqref{eq:Sbasis}. The eigenstates of the total-spin operator are
\begin{subequations}
    \begin{align}
        &\ket{1,1} \equiv \Sigma^+ = \Sigma^{++}, \\
        &\ket{1,-1} \equiv \Sigma^- = \Sigma^{--}, \\
        &\ket{1,0}\equiv \Sigma^{\bar{0}} = \frac{\Sigma^{+-}
+ \Sigma^{-+} }{\sqrt{2}}, \quad \text{and} \\
        &\ket{0} \equiv \Sigma^0= \frac{\Sigma^{+-}
- \Sigma^{-+} }{\sqrt{2}}.
    \end{align}\label{eq:ESSB}
\end{subequations}

Explicitly,
\begin{equation}
    \begin{split}
    & \ket{1,\pm 1}=
    \frac{1}{2}\left[\begin{pmatrix}
        1 & 0\\ 0 & 1
    \end{pmatrix} \pm\begin{pmatrix}
        1 & 0\\ 0 & -1
\end{pmatrix} \right], \\ & 
    \ket{1,0}=
    \frac{1}{\sqrt{2}}\begin{pmatrix}
        0 & 1 \\ 1 & 0
    \end{pmatrix} \quad \text{and} \quad \ket{0}=
    \frac{1}{\sqrt{2}}
    \begin{pmatrix}
        0 & 1 \\ -1 & 0
    \end{pmatrix}.
\end{split}
\end{equation}

In the matrix representation, we immediately observe that
$\sqrt{2}\ket{1,0}$ is represented by $\sigma_1$. Similarly, depending on the
phase convention chosen for the singlet state, $\sqrt{2}\ket{0}$ is represented
by $\pm i\sigma_2$. With the convention used here, this gives
$\sqrt{2}\ket{0}=i\sigma_2$. This observation motivates the use of the basis
$\{\mathds{1},\bm{\sigma}\}$, which is particularly convenient because the
spin operators are proportional to the Pauli matrices. In this basis, the
states become
\begin{subequations}
    \begin{align}
        &\ket{1,1}   =  \frac{\mathds{1}+\sigma_3}{2}, \\
        &\ket{1,-1}   =  \frac{\mathds{1}-\sigma_3}{2}, \\
        &\ket{1,0} =  \frac{\sigma_1}{\sqrt{2}},  \quad \text{and} \\
        &\ket{0}   =  i\frac{\sigma_2}{\sqrt{2}}.
    \end{align}\label{eq:ESPB}
\end{subequations}

Returning to the expectation value~\eqref{eq:bellexp},
\begin{align} \label{eq:bell}
    B \left [ {\psi}\right ]&=\ev**{S^{(1)}_{\hat{\bm{u}}} S^{(2)}_{\hat{\bm{v}}}}{\psi}
    \nonumber \\
    &= u_j{v}_k\ev**{S_j^{(1)}S_k^{(2)}}{\psi}
    \nonumber \\
    &=
    \frac{\hbar^2}{4}u_j{v}_k
    K_{jk}[\psi],
\end{align}
we define
\begin{equation}
    K_{jk} \left [ {\psi} \right ] \equiv
    \ev**{\sigma_j^{(1)}\sigma_k^{(2)}}{\psi}.
    \label{eq:Kspin}
\end{equation}

Using the inner product~\eqref{eq:sprod}, $K_{jk} [ \psi ]$ can be expressed as 
\begin{equation}
    \Tr[ (\Sigma^{\psi}  ){}^\dagger
    \sigma_j^{(1)}\sigma^{(2)}_k\Sigma^{\psi}],
    \label{eq:Kspin1}
\end{equation}
where $\Sigma^{\psi}$ is the matrix representing the c-state $\psi$. Using equation~\eqref{eq:main}, we obtain 
\begin{equation}
    K_{jk} \left [ {\psi} \right ] =
    \Tr[ (\Sigma^{\psi}  ){}^\dagger
    \sigma_j\Sigma^{\psi}\sigma_k^{\mathsf{T}}].
    \label{eq:main2}
\end{equation}

In general, for any operator $A\otimes B \in \mathcal{L}(\mathcal{\hat H})$, we have 
\begin{equation}
    \ev**{A \otimes B}{\psi}=
    \Tr[ (\Sigma^{\psi} ){}^\dagger A\Sigma^{\psi}B^{\mathsf{T}}].
    \label{eq:GEV1}
\end{equation}

\subsubsection{Singlet}\label{sec:singlet}

We start with the singlet state. Since we already know the answer, it is
straightforward to anticipate that $K$ must be equal to $-\delta_{jk}$ for the
singlet, that is, when $\ket{\psi}=\ket{0}$. However, our interest here lies in
the method leading to this result. Once equation~\eqref{eq:main2} has been obtained,
the calculation becomes almost immediate. Replacing $\ket{\psi}$ with the
singlet state we have $\Sigma^{0}=i\sigma_2/\sqrt{2}$, so that
equation~\eqref{eq:main2} becomes
\begin{equation}
    2K_{jk} \left [ {0} \right ] =
    \Tr[ \sigma_2 \sigma_j\sigma_2\sigma_k^{\mathsf{T}}],
    \label{eq:singlet1}
\end{equation}
where the Hermiticity of the Pauli matrices has been used.
We will also employ other well-known properties of the Pauli matrices.
For clarity, they are listed in the Appendix.

Let us begin with $\sigma_2\sigma_j\sigma_2$. Applying Property~2 several
times, we get
\begin{align}
    \sigma_2\sigma_j\sigma_2 &= \sigma_2(\delta_{2j}\mathds{1} +i\epsilon_{j2l}
    \sigma_l) \nonumber \\
    &= \sigma_2 \delta_{2j} +i\epsilon_{j2l}(\delta_{2l}\mathds{1}+i
    \epsilon_{2ln}\sigma_n) \nonumber \\
    &=\sigma_2 \delta_{j2} +\epsilon_{j2l}\epsilon_{n2l}\sigma_n
    \nonumber \\
    &=\sigma_2\delta_{j2} +(\delta_{j2}\delta_{2n}-\delta_{22}\delta_{jn})
    \sigma_n \nonumber \\
    &= 2\sigma_2 \delta_{j2} - \sigma_j,
    \label{eq:sigmaP}
\end{align}
where in the penultimate line we used
\begin{equation}
    \epsilon_{jkl}\epsilon_{mnl}=\delta_{jm}\delta_{kn}-\delta_{jn}\delta_{km}.
    \label{eq:propEPS}
\end{equation}

We could substitute this result directly into equation~\eqref{eq:singlet1}, but
there is a more convenient way. Listing all cases of the last equality yields
\begin{equation}
      \sigma_2\sigma_j\sigma_2 =
      \begin{cases}
          \sigma_2,  &\text{if } j=2,\\
          -\sigma_j, &\text{if } j\neq2,
      \end{cases}
\end{equation}
which is equal to the transpose of $-\sigma_j$ (see Property~5), that is,
\begin{equation}
      \sigma_2\sigma_j\sigma_2 = -\sigma_j^\mathsf{T}.
      \label{eq:singlet2}
\end{equation}

Combining this result with equation~\eqref{eq:singlet1} we obtain
\begin{equation}
    \Tr[ \sigma_2
    \sigma_j\sigma_2\sigma_k^{\mathsf{T}}] =-\Tr[\sigma_j^\mathsf{T}
    \sigma_k^\mathsf{T}]= -\Tr[\sigma_k\sigma_j],
    \label{eq:singlet3}
\end{equation}
and using Property~2 once more we find
\begin{equation}
    -\Tr[\sigma_k\sigma_j]=-\Tr[\mathds{1}\delta_{jk}+i\epsilon_{kjl}\sigma_l]
    =-2\delta_{jk},
    \label{eq:singlet4}
\end{equation}
since the Pauli matrices are traceless. Therefore,
\begin{equation}
    K_{jk}\left [ {0}\right ] = -\delta_{jk}.
    \label{eq:bell1}
\end{equation}

Substituting this result into equation~\eqref{eq:bell} we obtain the expectation
value for the singlet:
\begin{align}
    B \left [0\right ] &=\frac{\hbar^2}{4}u_j{v}_k
\ev**{\sigma_j^{(1)}\sigma_k^{(2)}}{0} \nonumber \\
&=-\frac{\hbar^2}{4} u_{j}v_k\delta_{jk} \nonumber \\
&= -\frac{\hbar^2}{4}\hat{\bm{u}}\cdot\hat{\bm{v}} = -\frac{\hbar^2}{4}\cos{\theta},
\end{align}
which agrees with our earlier result ~\eqref{eq:bellS0}.

Equation~\eqref{eq:singlet4} shows that the Pauli matrices are orthogonal.
Denoting the identity matrix by $\sigma_0$, that is
$\sigma_{0} \equiv \mathds{1}$, it is well known that
\begin{equation}
    \Tr[\sigma_{\alpha}\sigma_\beta]=2\delta_{\alpha\beta},
\end{equation}
where, as in General Relativity, the Greek indices run from $0$ to $3$.
Therefore the basis $\{\mathds{1},\,\bm{\sigma}\}$ is orthogonal.

\subsubsection{Triplet}

We now turn to the triplet states. We begin with $\ket{\psi}
=\ket{1,0}$, since it is similar to the singlet case. In this case
$\Sigma^{\bar{0}} = \sigma_1/\sqrt{2}$, hence
\begin{equation}
    2K_{jk}\left [ 1, 0\right ] =\Tr[\sigma_1\sigma_j\sigma_1
    \sigma_k^\mathsf{T}].
    \label{eq:tripletS0}
\end{equation}

Following the same strategy as before, we first compute
$\sigma_1\sigma_j\sigma_1$. One can easily verify that
\begin{equation}
    \sigma_1\sigma_j\sigma_1 = 2\sigma_1\delta_{j1}-\sigma_j,
    \label{eq:tripletS1}
\end{equation}
which is analogous to equation~\eqref{eq:sigmaP}, and its demonstration is
similar. A similar relation holds for $\sigma_3\sigma_j\sigma_3$ after
replacing $1$ with $3$.

Substituting equation~\eqref{eq:tripletS1} into $K$ yields
\begin{equation}
    2K_{jk}\left [ 1, 0\right ] =2\delta_{j1}\Tr[\sigma_1
    \sigma_k^\mathsf{T}]-\Tr[\sigma_j\sigma_k^\mathsf{T}].
    \label{eq:tripletS2}
\end{equation}

Since $\sigma_1$ is symmetric, the first term follows from the orthogonality
of the Pauli matrices,
$\Tr[\sigma_1\sigma_k]=2\delta_{k1}$, and therefore the first term equals
$4\delta_{j1}\delta_{k1}$. For the second term we use Property~5,
\begin{equation}
    -\Tr[\sigma_j\sigma_k^\mathsf{T}]=
    \begin{cases}
    2\delta_{j2},&\text{when } j=2; \\
        -2\delta_{jk},&\text{when } j\neq2.
    \end{cases}
\label{eq:prop1}
\end{equation}

Thus,
\begin{equation}
    K_{jk} \left [ 1, 0\right ] =\delta_{j1}\delta_{k1}
    +\delta_{j2}\delta_{k2} -\delta_{j3}\delta_{k3}.
    \label{eq:tripletS3}
\end{equation}

Substituting this into $B$ [see equation~\eqref{eq:bell}] gives
\begin{align}
    B \left [ 1, 0 \right ] &=\frac{\hbar^2}{4}u_j{v}_k
    \left (\delta_{j1}\delta_{k1}
    +\delta_{j2}\delta_{k2} -\delta_{j3}\delta_{k3}
\right ) \nonumber \\
&= \frac{\hbar^2}{4}\left (u_1v_1+u_2v_2-u_3v_3\right ) \nonumber \\
&= \frac{\hbar^2}{4}\left (\sin \theta_u \sin \theta_v
        \cos{(\varphi_u-\varphi_v)}-\cos \theta_u \cos \theta_v \right ).
\end{align}
As expected, this agrees with the value obtained using the first approach shown in equation~\eqref{eq:bellSP} [in which, we have used equation~\eqref{eq:angletheta}].

We now proceed to the states $\ket{1,1}$ and $\ket{1,-1}$. We treat them
together by setting $\ket{\psi} =\ket{1,\pm1}$ and carrying the double signs
($\pm$ and $\mp$) throughout the calculation. The matrices representing these states can be written compactly as
$\Sigma^{\pm} =(\mathds{1} \pm \sigma_3)/2$. Thus,
\begin{equation}
    4K_{jk} [1, \pm 1] =\Tr[(\mathds{1}\pm\sigma_3)\sigma_j
    (\mathds{1}\pm\sigma_3) \sigma_k^\mathsf{T}].
    \label{eq:tripletpm0}
\end{equation}

Following the previous procedure, we compute
$(\mathds{1}\pm\sigma_3)\sigma_j (\mathds{1}\pm\sigma_3)$ first. Using
Property~2 repeatedly, the expression reduces to a linear combination of
$\{\mathds{1},\,\bm{\sigma}\}$. In general, the product of Pauli matrices
(or any linear combination of them) remains an element of the linear space
$\mathbb{C}^{2\times 2}$ and can therefore be expressed in the basis
$\{\sigma_\alpha\}$.

Thus,
\begin{align}
    (\mathds{1}\pm\sigma_3)\sigma_j (\mathds{1}\pm\sigma_3)   &=
    (\sigma_j\pm\sigma_3\sigma_j)(\mathds{1}\pm\sigma_3) \nonumber \\
    &=\sigma_j\pm\sigma_j\sigma_3 \pm \sigma_3\sigma_j +\sigma_3\sigma_j
    \sigma_3 \nonumber \\
    &= \sigma_j+\sigma_3\sigma_j\sigma_3 \pm \{\sigma_3,\,\sigma_j\} \nonumber \\
    &= 2(\sigma_3\delta_{j3}\pm \delta_{3j}\mathds{1}),
\label{eq:tripletpm1}
\end{align}
where the anticommutator $\{\sigma_3, \, \sigma_j \}$ equals
$2\delta_{j3}\mathds{1}$ and
$\sigma_3\sigma_j\sigma_3=2\sigma_3\delta_{j3}-\sigma_j$.

Substituting equation~\eqref{eq:tripletpm1} into equation~\eqref{eq:tripletpm0} yields
\begin{equation}
    2K_{jk} [1, \pm 1]
    =\delta_{j3}\Tr[\sigma_3\sigma_k^\mathsf{T}]
    \pm\Tr[\sigma_k^\mathsf{T}],
    \label{eq:tripletpm2}
\end{equation}
and the second term vanishes because Pauli matrices are traceless. Using
equation~\eqref{eq:prop1} with $j=3$, we obtain
\begin{equation}
    K_{jk}[1, \pm 1]
    =\delta_{j3}\delta_{k3}.
\label{eq:tripletpm3}
\end{equation}

Substituting this result into equation~\eqref{eq:bell} gives
\begin{align}
    B [1, \pm 1] &=\frac{\hbar^2}{4}u_{j} v_{k}\delta_{j3}\delta_{k3}
    \nonumber \\
    &=\frac{\hbar^2}{4}\cos{\theta_u}\cos{\theta_v},
\end{align}
which coincides with the result obtained using the first approach equations~\eqref{eq:bellMR} and~\eqref{eq:bellMN}.

This second approach is more formal than the product-basis calculation,
but it provides a clearer and more efficient route to the spin correlations.
Once its basic ingredients are established, in particular the action of
operators in equation~\eqref{eq:main} and the inner product in
Eq.~\eqref{eq:sprod}, the evaluation of the correlation function becomes
considerably simpler. This is one of the main pedagogical advantages of the
method: it reorganizes the calculation in terms of familiar $2\times2$ matrix
operations, making the tensor-product structure of the bipartite system more
transparent and reducing the amount of algebra needed to reach the final
results.

In comparison with the direct approach, the matrix representation reduces the
amount of algebra required and avoids several trigonometric substitutions that
can obscure the physical meaning of the result. The required identities are
either standard consequences of the tensor-product structure or can be derived
in a few steps once a basis is chosen. Thus, while the method requires a short
preliminary discussion, it offers a more compact and conceptually organized
derivation of the correlation function.

\section{Symmetry-based approach}
\label{sec:symmetry}

The third approach is based on symmetry considerations. It is inspired by
the argument used by Griffiths in Problem 4.50 of
\textit{Introduction to Quantum Mechanics} \cite{G1}, where the spin
correlation is evaluated for the singlet state.

The main idea is to exploit the freedom to choose convenient axes. For the
singlet, one may choose $\hat{\bm{u}}$ along $\hat{\bm{k}}$, so that
$S_{\hat{\bm{u}}}=S_z$, and take $\hat{\bm{v}}$ in the $xz$-plane, namely
\[
S_{\hat{\bm{v}}}=\sin\theta\,S_x+\cos\theta\,S_z,
\]
where $\theta$ is the angle between $\hat{\bm{u}}$ and $\hat{\bm{v}}$. This procedure gives
the correct result for the singlet,
\[
B[0]=-\frac{\hbar^2}{4}\cos\theta.
\]
The reason is that the singlet is rotationally invariant. Therefore, rotating
the measurement directions does not change the state.

However, a common pitfall is to extend this same reasoning directly to the
triplet states. If one applies the same shortcut to the triplet sector, one
finds
\begin{equation}
    B [1, \pm 1] = -B \left [ 1,0 \right ]
    = \frac{\hbar^2}{4}\cos\theta .
\end{equation}
This result is rotationally invariant, since it depends only on the relative
angle between $\hat{\bm{u}}$ and $\hat{\bm{v}}$. It therefore disagrees with the results
obtained by the direct and matrix-based methods.

The failure of the naive symmetry argument can be seen already for the state
$\ket{1,1}$. The correct result is
\begin{equation}
B[1,1]=\frac{\hbar^2}{4}\cos\theta_u\cos\theta_v .
\label{eq:B11correct}
\end{equation}
If one chooses
\[
\hat{\bm{u}}=\hat{\bm{k}},\qquad
\hat{\bm{v}}=\sin\theta\,\hat{\bm{\imath}}+\cos\theta\,\hat{\bm{k}} ,
\]
then equation~\eqref{eq:B11correct} gives
\[
B[1,1]=\frac{\hbar^2}{4}\cos\theta .
\]
A student might then incorrectly conclude that the triplet correlation depends
only on the relative angle between the two measurement directions. This is not
the case. Consider instead
\[
\hat{\bm{u}}=\hat{\bm{\imath}},\qquad
\hat{\bm{v}}=\cos\theta\,\hat{\bm{\imath}}+\sin\theta\,\hat{\bm{k}} .
\]
The relative angle between $\hat{\bm{u}}$ and $\hat{\bm{v}}$ is again $\theta$, but now
$\cos\theta_u=0$, and therefore
\[
B[1,1]=0 .
\]
Thus, two configurations with the same relative angle lead to different
correlations. This counterexample shows where the naive symmetry reasoning
fails: it treats the triplet state as if it were rotationally invariant.

The origin of the disagreement is therefore not a failure of symmetry itself,
but an incomplete use of it. When the axes are rotated, the state must also be
transformed. For the singlet this point is hidden, because
$\ket{0}$ is rotationally invariant. For triplet states, however, the state changes
under the same rotation. Thus, rotating the measurement directions while
keeping the triplet state fixed changes the physical problem.

Next, we will see that to apply the symmetry-based approach consistently, one must rotate both the
operators and the state. For the singlet, $\ket{\psi'}=\ket{0}$, so the shortcut works immediately.
For triplet states, $\ket{\psi'}$ is generally a linear combination of triplet
states, and this transformation must be taken into account.

\subsection{Extending the symmetry approach}
It is not difficult to understand the origin of this disagreement. We restate
the argument in a clearer form. To choose the axes while keeping an equivalent
system, both subsystems must be rotated in the same way. Therefore, if we set
$\hat{\bm{u}} = \hat{\bm{k}}$ and
$\hat{\bm{v}}=\sin \theta \hat{\bm{\imath}} + \cos \theta \hat{\bm{k}}$,
we must find a unitary operator $U$ such that
\begin{equation}\label{eq:Ucond}
    US_{\hat{\bm{u}}}U^\dagger=S_z \quad \text{and} \quad
    US_{\hat{\bm{v}}}U^\dagger={\cos \theta}S_z+ {\sin{\theta}}S_x,
\end{equation}
that is,
$ \hat{U} S_{\hat{\bm{u}}}\otimes S_{\hat{\bm{v}}} \hat{U}^\dagger
= S_z \otimes \left ({\cos \theta}S_z
+ {\sin{\theta}}S_x \right )$,
where $U \in SU(2)$. Since the operators transform in this manner while
the spin operators themselves remain unchanged, the basis must also transform.
A state $\ket{a}\otimes\ket{b}$ must transform as
$\left (\ket{a}\otimes \ket{b} \right)' \equiv
\hat{U}\ket{a} \otimes \ket{b}$ (remind that $ \hat{U} = U \otimes U$)
in order to preserve the inner product.

Hence the disagreement arises because we were not solving the same problem.
The obtained values are not the general quantity $B$, but correspond to the
special case $\theta_{u}=0$. Nevertheless, this approach gives the correct
result for the singlet. This immediately suggests the existence of $U$ and
also indicates that the singlet is invariant under such rotations,
namely $U\otimes U\ket{0}=\ket{0}$.
Instead of proving invariance under a specific $U$, we prove the stronger
statement that the singlet is rotationally invariant.

\begin{proof}
Let
\begin{equation}\label{eq:defD}
   D =
   \begin{pmatrix}
       D_{11} & D_{12} \\
       D_{21} & D_{22}
   \end{pmatrix},
\end{equation}
be an element of $SL(2,\mathbb{C})$, the group of all $2\times2$ matrices
with unit determinant under matrix multiplication and inversion.
In physics, $D$ is usually taken to be unitary and therefore restricted to
the subgroup $SU(2)$.\footnote{Since $SU(2)$ is a compact connected Lie group,
it represents spin rotations in quantum mechanics. (see Ref.~\cite{nakahara, georgi} for a discussion of group theory.}

Although it suffices to consider unitary transformations, this restriction
does not simplify the proof. The transformed basis vectors are
\begin{equation}
    D\chi_+=
    \begin{pmatrix}
        D_{11} \\ D_{21}
    \end{pmatrix}
    \quad \text{and} \quad
    D\chi_-=
    \begin{pmatrix}
        D_{12} \\ D_{22}
    \end{pmatrix},
\end{equation}
and therefore
\begin{equation}
    \sqrt{2} \ket{0}'=
    \begin{pmatrix}
        D_{11} \\ D_{21}
    \end{pmatrix}
    \otimes
    \begin{pmatrix}
        D_{12} \\ D_{22}
    \end{pmatrix}
    -
    \begin{pmatrix}
        D_{12} \\ D_{22}
    \end{pmatrix}
    \otimes
    \begin{pmatrix}
        D_{11} \\ D_{21}
    \end{pmatrix}.
\end{equation}

Using the matrix representation, we have 
\begin{align}
     (\Sigma^0 )'&= \frac{1}{\sqrt{2}}
   \begin{pmatrix}
       D_{11}D_{12}-D_{12}D_{11} & D_{11}D_{22} -D_{12}D_{21}\\
       D_{21}D_{12}-D_{22}D_{11}&  D_{21}D_{22}-D_{22}D_{21}
   \end{pmatrix} \nonumber \\
   &=\det(D)\Sigma^0.
\end{align}
Since $\det(D)=1$ for $SL(2,\mathbb{C})$,
\begin{equation}
     (\Sigma^0 )' = \Sigma^0
    \iff \ket{0}'=\ket{0}.
\end{equation}
\end{proof}

This explains why the symmetry argument yields the correct value for $B[0]$
and why $B[0]$ depends only on the angle between
$ \hat{\bm{u}} $ and $ \hat{\bm{v}} $.

Repeating the same procedure for the triplet gives
\begin{subequations}
\begin{align}
    (\Sigma^{+}  )' ={}& 
    \begin{pmatrix}
        D_{11}^{2} & D_{11}D_{21} \\
        D_{11}D_{21} & D_{21}^{2}
    \end{pmatrix}
    \nonumber\\
    ={}& D_{11}^{2} \Sigma^{+}
    + D_{11}D_{21}\sqrt{2}\Sigma^{\bar{0} }
    + D_{21}^{2}\Sigma^{-}, \\
    ( \Sigma^{-} )' ={}&
    \begin{pmatrix}
        D_{12}^{2} & D_{12}D_{22} \\
        D_{12}D_{22} & D_{22}^{2}
    \end{pmatrix}
    \nonumber\\
    ={}& D_{22}^{2} \Sigma^{-}
    + D_{12}D_{22}\sqrt{2}\Sigma^{\bar{0}}
    + D_{12}^{2} \Sigma^{+}, \\
       ( \Sigma^{\bar{0} }  )' ={}&
    \frac{1}{\sqrt{2}}
    \begin{pmatrix}
       D_{11}D_{12}+D_{12}D_{11} &
       D_{11}D_{22}+D_{12}D_{21}\\
       D_{21}D_{12}+D_{22}D_{11} &
       D_{21}D_{22}+D_{22}D_{21}
   \end{pmatrix}
   \nonumber \\
    ={}&\ \sqrt{2}D_{11}D_{12}\Sigma^{+}
   +(D_{11}D_{22}+D_{12}D_{21})\Sigma^{\bar{0}}
    \nonumber\\ & +\sqrt{2}D_{21}D_{22}\Sigma^{-}.
\end{align}
\end{subequations}

These expressions show that the triplet is not rotationally invariant.
However, if $ D = \exp(-i \varphi \sigma_{3}/2) $ represents a rotation
about the $z$-axis, then
\begin{subequations}\label{eq:transforz}
    \begin{align} 
    \ket{1,1}' &= e^{i\varphi}\ket{1,1}, \\
    \ket{1,-1}' &= e^{-i\varphi}\ket{1,-1}, \\
    \ket{1,0}' &= \ket{1,0}.
\end{align}
\end{subequations}
Hence the triplet is physically invariant under rotations about the
$z$-axis. Since both $\ket{0}$ and $\ket{1,0}$ remain unchanged under
such rotations, any superposition of them is also invariant, whereas
superpositions of $\ket{1,1}$ and $\ket{1,-1}$ generally acquire a physically
relevant relative phase.

We now determine $U$ satisfying the condition~\eqref{eq:Ucond}.
Using the relation $SO(3)\cong SU(2)/Z_2$ \cite{georgi}, we have
$US_jU^\dagger=\mathcal{U}_{jk}S_k$, where $\mathcal{U}$ is an $SO(3)$
rotation matrix. Let
$D_{\bm{w}}(\varphi)=\exp(-i\varphi\bm{w}\cdot S)$ denote a spin rotation.
Then
\begin{equation}
    D_{\bm{w}}(\varphi) S_{j} D_{\bm{w}}^{\dagger}(\varphi)
    =
    \left[R_{\bm{w}}(\varphi)\right]_{jk}S_k,
\end{equation}
where $R_{\bm{w}}(\varphi)$ is the corresponding $SO(3)$ rotation matrix.

For $U_{z} ( \varphi )=\exp(-i\varphi\sigma_{3}/2)$,
\begin{equation}
\label{eq:Uforz}
\mathcal{U}=R_z(\varphi)=
\begin{pmatrix}
\cos\varphi & \sin\varphi & 0\\
-\sin\varphi & \cos\varphi & 0\\
0 & 0 & 1
\end{pmatrix},
\end{equation}
while for $\exp(-i\varphi\sigma_{2}/2)$, we have
\begin{equation}
\label{eq:Ufory}
R_y(\varphi)=
\begin{pmatrix}
\cos\varphi & 0 & -\sin\varphi\\
0 & 1 & 0\\
\sin\varphi & 0 & \cos\varphi
\end{pmatrix}.
\end{equation}

If a rotation $\mathcal{U}$ exists such that
\begin{equation}\label{eq:cond1}
\mathcal{U}\hat{\bm{u}}=\hat{\bm{k}},\qquad
\mathcal{U}\hat{\bm{v}}=\cos\theta\,\hat{\bm{k}}+\sin\theta\,\hat{\bm{\imath}},
\end{equation}
then the corresponding $U$ exists. Such $\mathcal{U}$ is obtained as a
composition of rotations. Using Euler angles,
$\mathcal{U}=R_z(\gamma)R_y(\beta)R_z(\alpha)$, with
$\alpha=\varphi_u$, $\beta=\theta_u$, and
$\tan\gamma=\mathcal{I}/\mathcal{R}$, where
\begin{align}
\mathcal{R}&=\cos\theta_u\sin\theta_v
\cos(\varphi_v-\varphi_u)-\sin\theta_u\cos\theta_v,\\
\mathcal{I}&=\sin\theta_v\sin(\varphi_v-\varphi_u).
\end{align}
The useful relations are
\begin{align}
    \mathcal{R}=\cos\gamma\sin\theta \quad \text{and} \quad 
\mathcal{I}=\sin\gamma\sin\theta,
\end{align}
implying $\sin\theta=\sqrt{\mathcal{R}^2+\mathcal{I}^2}$.

Hence
\begin{equation}
R_z(\gamma)=
\csc\theta
\begin{pmatrix}
\mathcal{R} & \mathcal{I} & 0\\
-\mathcal{I} & \mathcal{R} & 0\\
0 & 0 & \sin\theta
\end{pmatrix},
\end{equation}
and
\begin{equation}
U=D_z(\gamma)D_y(\theta_u)D_z(\varphi_u).
\end{equation}

The singlet is the only invariant eigenstate,
\begin{equation}
\ket{0}'=\ket{0},
\end{equation}
while the triplet transforms as
\begin{subequations}
\begin{align}
\ket{1,1}' &= U_{11}^{2}\ket{1,1}
+U_{11}U_{21}\sqrt{2}\ket{1,0}
+U_{21}^{2}\ket{1,-1},\\
\ket{1,-1}' &= U_{22}^{2}\ket{1,-1}
+U_{12}U_{22}\sqrt{2}\ket{1,0}
+U_{12}^{2}\ket{1,1},\\
\ket{1,0}' &=
\sqrt{2}U_{11}U_{12}\ket{1,1}
+(U_{11}U_{22}+U_{12}U_{21})\ket{1,0}
\nonumber\\
           &\quad +\sqrt{2}U_{21}U_{22}\ket{1,-1}.
\end{align}
\label{eq:trans1}
\end{subequations}
This explains why the symmetry approach fails for the triplet but succeeds
for the singlet. 

Moreover, for a general state $\ket{\psi}$, we have
\begin{align}
B[\psi]
&=\ev**{S^{(1)}_{\hat{\bm{u}}}S^{(2)}_{\hat{\bm{v}}}}{\psi}\nonumber\\
&=\ev**{S_z^{(1)}
\left(\sin\theta S_x^{(2)}+\cos\theta S_z^{(2)}\right)}{\psi'},
\label{eq:BGG}
\end{align}
where $\ket{\psi'}= \hat{U} \ket{\psi}$. Since $\ket{\psi}$ can be decomposed in the eigenstate basis\footnote{For a normalized state $ |c_{0} |^{2} + |c_{\bar{0}} |^{2} + | c_{-} |^{2} + |c_{+}|^{2} =1 $.}
\begin{equation}\label{eq:transpsi}
    \ket{\psi} = c_{0} |0\rangle + c_{\bar{0}} |1,0\rangle + c_{-} |1,-1\rangle + c_{+} |1,1\rangle, 
\end{equation}
its transformation follows as 
\begin{equation}\label{eq:transpsi2}
    \ket{\psi'} = c_{0} |0\rangle + c_{\bar{0}} |1,0\rangle' + c_{-} |1,-1\rangle' + c_{+} |1,1\rangle', 
\end{equation}
where the primed states are shown in equation~\eqref{eq:trans1}.

A limited use of symmetry is still possible in the triplet sector. Consider,
for example, the states $\ket{1,\pm1}$ and the special case in which one of
the measurement directions is the $z$-axis,
\begin{equation}
    B_{\hat{\bm{k}},\hat{\bm v}}[1,\pm1]
    =
    \ev**{S^{(1)}_{\bm{k}}S^{(2)}_{\hat{\bm v}}}{1,\pm1}.
\end{equation}
Writing
\begin{equation}
    S^{(2)}_{\hat{\bm v}}
    =
    \sin\theta_v\cos\varphi_v\,S_x^{(2)}
    +
    \sin\theta_v\sin\varphi_v\,S_y^{(2)}
    +
    \cos\theta_v\,S_z^{(2)},
\end{equation}
one may use the fact that $\ket{1,\pm1}$ are eigenstates of the total
$S_z$ operator. Under a rotation around the $z$-axis, these states acquire
only an overall phase [see equation~\eqref{eq:transforz}], which cancels in the expectation value. Therefore, one may choose the azimuthal angle of
$\hat{\bm v}$ to be zero\footnote{This is possible by choosing $U_{z} = \exp(i \varphi_{v} \sigma_{3} /2)$, which corresponds to a rotation of angle $ - \varphi_{v} $ around the $z$-axis.}, so that
\begin{equation}
    S^{(2)}_{\hat{\bm v}}
    \longrightarrow
    \sin\theta_v\,S_x^{(2)}
    +
    \cos\theta_v\,S_z^{(2)}.
\end{equation}
Thus,
\begin{align}
    B_{\hat{\bm{k}},\hat{\bm v}}[1,\pm1]
    &=
    \ev**{
    S_z^{(1)}
    \left(
    \sin\theta_v\,S_x^{(2)}
    +
    \cos\theta_v\,S_z^{(2)}
    \right)
    }{1,\pm1}  \nonumber \\
    &=
    \frac{\hbar^2}{4}\cos\theta_v .
\end{align}
This agrees with Eqs.~\eqref{eq:bellMR} and~\eqref{eq:bellMN} for
$\theta_u=0$, namely $\hat{\bm u}=\hat{\bm{k}}$.

It is important, however, that this is only a special use of symmetry. It works
because $\ket{1,\pm1}$ transform by an overall phase under rotations around
the $z$-axis [see equation~\eqref{eq:transforz}]. For a superposition such as
$\ket{1,1}+\ket{1,-1}$, the two components acquire different phases under the
same rotation. The relative phase of the state is therefore changed, and the
above simplification can no longer be applied without also transforming the
state explicitly.

This shows that extending Griffiths's approach to other states requires transforming it according to equation~\eqref{eq:transpsi2}, so that $B$ remains invariant under the transformation of $S_{\hat{\bm{u}}}$ and $S_{\hat{\bm{v}}}$, as in equation~\eqref{eq:BGG}.

\section{Conclusion}

In this work we presented a pedagogical discussion of spin correlations in a
two--particle spin-$1/2$ quantum system, emphasizing different strategies for
evaluating expectation values of the form
$\ev{S^{(1)}_{\hat{\bm{u}}} S^{(2)}_{\hat{\bm{v}}}}{\psi}$.
Rather than introducing new physical results, the objective was to compare
distinct derivational approaches and clarify their conceptual content.

The direct calculation in the product basis provides an explicit and accessible
procedure, although it may become algebraically lengthy. The formulation based
on a matrix representation of bipartite states leads to a more concise
treatment and makes transparent the independent action of operators on each
subsystem. In contrast, the symmetry argument inspired by Griffiths,
while successful for the singlet state, was shown to fail when naively extended
to the triplet sector. This difference was traced to the rotational invariance
of the singlet state and to the transformation properties of the triplet
states under spin rotations.

From a pedagogical perspective, the comparison between these approaches helps
to connect algebraic manipulation, geometric symmetry, and physical
interpretation. In particular, the analysis clarifies a common source of
difficulty for students: the implicit assumption that symmetry arguments
apply equally to all states. By exhibiting explicitly the failure of this
reasoning in the triplet sector, the discussion highlights the importance of
distinguishing between transformations of measurement axes and transformations
of the quantum state.

Moreover, the matrix-based representation provides a more transparent view of
the tensor-product structure, allowing the calculation to be organized in terms
of independent contributions from each subsystem. This can be especially useful
for students encountering bipartite systems for the first time, as it reduces
the reliance on lengthy algebraic manipulations and makes the structure of
entangled states more explicit.

Finally, the quantities studied here have a direct experimental interpretation.
The expectation value $\ev{S^{(1)}_{\hat{\bm{u}}} S^{(2)}_{\hat{\bm{v}}}}{\psi}$ corresponds
to the average product of spin measurements performed along directions
$\hat{\bm{u}}$ and $\hat{\bm{v}}$, as in Stern--Gerlach or Bell-type experiments. In this
sense, the different methods discussed in this work not only provide alternative
computational tools, but also offer complementary perspectives that connect
formal calculations with physically measurable correlations.

We hope that the discussion presented here may serve as a complementary resource
for undergraduate and introductory graduate courses in quantum mechanics,
especially in topics related to spin systems \cite{sakurai, G1}, quantum entanglement \cite{nielsen, ballentine}, and Bell-type correlations \cite{bell, clauser}.

The third method is particularly effective for the singlet, although less
useful for the triplet due to its reduced symmetry. For the singlet, it combines
simplicity and conciseness and therefore provides a useful introduction for
undergraduate students to the EPR paradox and Bell's inequalities, especially
in connection with Bell's original work \cite{bell}. In this respect,
the pedagogical presentation in Ref.~\cite{mermin1990} is also especially useful,
as it emphasizes the conceptual content of Bell-type correlations with minimal
formal machinery.

\begin{acknowledgments}

The author thanks Gastão Krein for his encouragement, which motivated the completion of this work.
The author also thanks Prof. Jeferson L. Tomazelli for useful comments and helpful suggestions.  
S.\ M.-F. thanks FAPESP for partial financial support. This study was financed, in part, by the São Paulo Research Foundation (FAPESP), Brasil. Process Number \#2025/16156-7. 
\end{acknowledgments}

\appendix

\section{Pauli Matrices}
Here we will list some of the properties of the Pauli matrices \cite{sakurai,ballentine}. These 
well-known properties were used extensively in this work, but specially in the 
Section~\ref{sec:second}: 
\begin{enumerate}[label=Property \arabic*.,itemindent=*]
\item 
Hermiticity:
$\sigma_j^\dagger = \sigma_j$.

\item 
Pauli algebra:
$\sigma_j\sigma_k = \mathds{1}\delta_{jk} + i
\epsilon_{jkl}\sigma_l$.

\item 
Involution:
$\sigma_j^2=\mathds{1}$.

\item 
Tracelessness:
$\Tr[\sigma_j]=0$.

\item 
Transposition:
$
\sigma_j^\mathsf{T}=
\begin{cases}
-\sigma_2, & \text{if } j=2,\\
+\sigma_j, & \text{if } j\neq2.
\end{cases}
$
\end{enumerate}

\end{document}